\documentclass[pra,twocolumn,10pt,aps,amsmath,amsfonts,superscriptaddress,floatfix]{revtex4-1}

\usepackage{stix}
\usepackage{graphicx}
\usepackage{dcolumn}
\usepackage{bm}
\usepackage{braket}
\usepackage{float}
\usepackage{amsmath}
\usepackage{hyperref}

\usepackage[usenames,dvipsnames]{color}

\newcommand{\W}{\mathbb{W}}
\renewcommand{\H}{\mathbb{H}}
\renewcommand{\t}{t}
\renewcommand{\tt}{\tau}
\newcommand{\R}{{\cal R}}
\newcommand{\J}{\mathbb{J}}
\newcommand{\RR}{\mathbb{R}}

\newcommand{\twobytwo}{\ensuremath{2\times 2}}

\makeatletter
\def\beq{\@ifstar{\@ifnextchar[{\@beqslabel}{\@beqsnolabel}}
{\@ifnextchar[{\@beqlabel}{\@beqnolabel}}}
\def\@beqlabel[#1]{\begin{equation}\label{#1}}
\def\@beqnolabel{\begin{equation}}
\def\@beqslabel[#1]{\begin{equation*}\label{#1}}
\def\@beqsnolabel{\begin{equation*}}
\def\eeq{\@ifstar{\end{equation*}}{\end{equation}}}
\makeatother

\newcommand{\punc}[1]{\;\text{#1}}
\newcommand{\refeq}[1]{Eq.~(\ref{#1})}
\newcommand{\refeqand}[2]{Eqs.~(\ref{#1}) and (\ref{#2})}
\newcommand{\reffig}[1]{Fig.~\ref{#1}}
\newcommand{\refsec}[1]{Section~\ref{#1}}
\newcommand{\refapp}[1]{Appendix~\ref{#1}}
\newcommand{\refcite}[1]{Ref.~\cite{#1}}

\newcommand{\vertdimer}{\textnormal{II}}
\newcommand{\horidimer}{\rotatebox{90}{\textnormal{II}}}

\newcommand{\ee}{e}
\newcommand{\ii}{\mathrm{i}}
\newcommand{\dd}{\mathrm{d}}

\newcommand{\rv}{\boldsymbol{r}}
\newcommand{\kv}{\boldsymbol{k}}
\newcommand{\Phiv}{\vec{\Phi}}
\newcommand{\deltav}{\boldsymbol{\delta}}
\newcommand{\Nv}{\vec{N}}
\newcommand{\zerov}{\vec{0}}
\newcommand{\goC}{\mathfrak{C}}
\newcommand{\goP}{\mathfrak{P}}
\newcommand{\scS}{\mathcal{S}}
\newcommand{\del}{\boldsymbol{\nabla}}
\newcommand{\sub}[1]{_{\text{#1}}}
\newcommand{\dsQ}{\mathbb{Q}}

\newcommand{\gs}{\text{gs}}
\newcommand{\SO}{\mathrm{SO}}

\begin{document}
	
\title{Phases of quantum dimers from ensembles of classical stochastic trajectories}

\author{Tom Oakes}\email{tom.oakes@nottingham.ac.uk}
\author{Stephen Powell}
\affiliation{School of Physics and Astronomy
and \\
Centre for the Mathematics and Theoretical Physics of Quantum Non-Equilibrium Systems, \\
University of Nottingham, Nottingham NG7 2RD, United Kingdom}

\author{Claudio Castelnovo}
\author{Austen Lamacraft}
\affiliation{TCM Group, Cavendish Laboratory, University of Cambridge, J. J. Thomson Avenue, Cambridge CB3 0HE, United Kingdom}

\author{Juan P. Garrahan}
\affiliation{School of Physics and Astronomy
and \\
Centre for the Mathematics and Theoretical Physics of Quantum Non-Equilibrium Systems, \\ 
University of Nottingham, Nottingham NG7 2RD, United Kingdom}

\date{\today}

\begin{abstract}
We study the connection between the phase behaviour of quantum dimers and the dynamics of classical stochastic dimers. At the so-called Rokhsar--Kivelson (RK) point a quantum dimer Hamiltonian is equivalent to the Markov generator of the dynamics of classical dimers. A less well understood fact is that away from the RK point the quantum--classical connection persists: in this case the Hamiltonian corresponds to a non-stochastic ``tilted'' operator that encodes the statistics of time-integrated observables of the classical stochastic problem. This implies a direct relation between the phase behaviour of  quantum dimers and properties of ensembles of stochastic trajectories of classical dimers. We make these ideas concrete by studying fully packed dimers on the square lattice. Using transition path sampling -- supplemented by trajectory umbrella sampling -- we obtain the large deviation statistics of dynamical activity in the classical problem, and show the correspondence between the phase behaviour of the classical and quantum systems. The transition at the RK point between quantum phases of distinct order corresponds, in the classical case, to a trajectory phase transition between active and inactive dynamical phases. Furthermore, from the structure of stochastic trajectories in the active dynamical phase we infer that the ground state of quantum dimers has columnar order to one side of the RK point. We discuss how these results relate to those from quantum Monte Carlo, and how our approach may generalise to other problems. 
\end{abstract}

\maketitle

\section{Introduction}

Dimer models are prototypical examples of systems where the degrees of freedom are subject to strong local constraints. Quantum dimer models (QDMs) were initially conceived by Anderson and collaborators \cite{Anderson1973,Fazekas1974} as representations of singlet pairings of quantum spins. The simplest QDMs \cite{RK1988} correspond to systems where  dimers are the basic degrees of freedom, they fully pack a lattice, and have a Hamiltonian with kinetic terms flipping groups of neighbouring dimers -- for example pairs of parallel dimers on the square lattice -- and potential terms counting the number of such flippable clusters. In spite of their apparent simplicity -- and of how old these models are -- we lack a full understanding the ground state phase behaviour of such QDMs \cite{Moessner2011,Chalker}. 

At zero temperature, the phase diagram of fully packed QDMs is controlled by the ratio $v/\t$ of 
the energy per plaquette $v$ to the flipping frequency $\t$ (see next section for definitions). The case $v/\t=1$, usually called the Rokhsar--Kivelson (RK) point, is of special importance. Here the ground state can be found exactly and is given by an equal superposition of all dimer configurations. The RK point often also delimits different ground state regimes. For example, for the square or honeycomb lattices, the ground state at the RK point is a spin liquid with algebraic decaying correlations (a ``Coulomb phase'' \cite{HenleyReview}), separating states with different kinds of order at either side of $v/\t=1$: for $v/\t>1$ this order is known to be ``staggered'', extending all the way to $v/\t = \infty$; for $v/\t<1$, in contrast, the nature of the order is less clear, the only certainty being that 
for $v/\t = -\infty$ it is ``columnar''  \cite{Moessner2011,Chalker}. The richness of this phase diagram highlights the complexity that can emerge from the simple ingredient of a constrained Hilbert space.

Constraints can also play a significant role in classical stochastic many-body systems. For example, kinetically constrained models (KCMs)  \cite{Ritort2003} are simple lattice systems with local constraints in their dynamics that mimic steric restictions, and are one of the paradigms for the slow relaxation characteristic of glassy systems \cite{Chandler2010}. Fully packed classical dimer models (CDMs) also have rich dynamical behaviour \cite{Henley1997,Oakes2016}.  And even the simplest constraint of hard-core repulsion as in simple exclusion processes can give rise to interesting non-equilibrium dynamics \cite{Derrida2007}. In order to fully capture the properties of such complex dynamics it is necessary to study the statistical properties of their trajectories. The natural framework is that of dynamical large deviations (LDs) \cite{Touchette2009}, which provides a {\em statistical mechanics of trajectories} \cite{DeformationPaper1,DeformationPaper2,DeformationPaper3} which is the dynamical analog of the standard equilibrium ensemble method for static configurations. In this approach, dynamical properties are classified according to time-integrated observables, and the long-time limit plays the role of the thermodynamic limit in the static case. In the long-time regime the statistics of dynamical observables is encoded in LD functions that play the role of dynamical entropies and free-energies. 

In this paper we aim to connect the static quantum phase behaviour of QDMs with the dynamical large deviation properties of the stochastic dynamics of CDMs. This connection starts with the quantum dimers at the RK point, where the Hamiltonian of the quantum system is the same as (minus) the generator of continuous time Markov dynamics of the classical dimers \cite{Henley2004,Laeuchli}. Away from the RK point the Hamiltonian is no longer a stochastic generator for the classical dynamics but is instead a deformation thereof, which corresponds to the {\em tilted} generator that encodes the statistics of the {\em dynamical activity} (number of configuration changes in a trajectory) of the classical system. This connects the statistical properties of classical trajectories to the spectral properties of the quantum problem. In what follows we establish these connections for the case of dimers on the square lattice.

\section{Model and definitions}

In both our classical and quantum models, the elementary degrees of freedom are dimers, which occupy the links of a lattice. We consider an \(L \times L\) square lattice with periodic boundary conditions and define \(d_\mu(\rv) \in \{0,1\}\) as the dimer occupation number on the link joining sites \(\rv\) and \(\rv + \deltav_\mu\), where \(\deltav_\mu\) is a lattice vector and \(\mu \in \{x,y\}\). The allowed configurations are those where the dimers are fully packed, i.e., where every site of the lattice is touched by precisely one dimer,
\beq
\sum_\mu \left[ d_\mu(\rv) + d_\mu(\rv - \deltav_\mu) \right] = 1\punc.
\eeq

\subsection{Quantum dimer model}
\label{SecQDM}

A complete basis for the Hilbert space of the QDM is given by all fully packed dimer configurations. We will use the RK Hamiltonian \cite{RK1988}, which can be written schematically as
\beq[QDMH]
\H_{\t,v} = - \t \sum(\ket{\vertdimer} \bra{\horidimer} + \ket{\horidimer} \bra{\vertdimer}) + v\sum(\ket{\vertdimer} \bra{\vertdimer} + \ket{\horidimer} \bra{\horidimer})\punc,
\eeq
where the sums are over the plaquettes (squares) of the lattice. The first summation is the kinetic energy, where each term flips a pair of dimers around a plaquette with frequency $\t$. The second summation is the potential energy which counts the number of flippable plaquettes, each plaquette carrying an energy $v$. 

Besides the total number of dimers, the Hamiltonian \(\H_{\t,v}\) in a lattice with periodic boundary conditions has a further conserved quantity, the {\it flux} \(\Phiv\). The components of the flux vector are defined by
\beq
\Phi_\mu = \sum_{\rv} \epsilon_{\rv} d_\mu(\rv)\punc,
\eeq
where \(\epsilon_{\rv} = (-1)^{r_x + r_y} = \pm 1\) on the two sublattices. 
The flux \(\Phiv\) is conserved by any local dynamics within the space of fully packed dimer configurations \cite{Moessner2011,Chalker}. Since there are \(\frac{1}{2}L^2\) dimers and each contributes \(\pm 1\) to one component of \(\Phiv\), possible values of the latter satisfy \(\lvert \Phi_x \rvert + \lvert \Phi_y \rvert \le \frac{1}{2}L^2\).

\begin{figure}
\begin{center}
\includegraphics[width=\columnwidth]{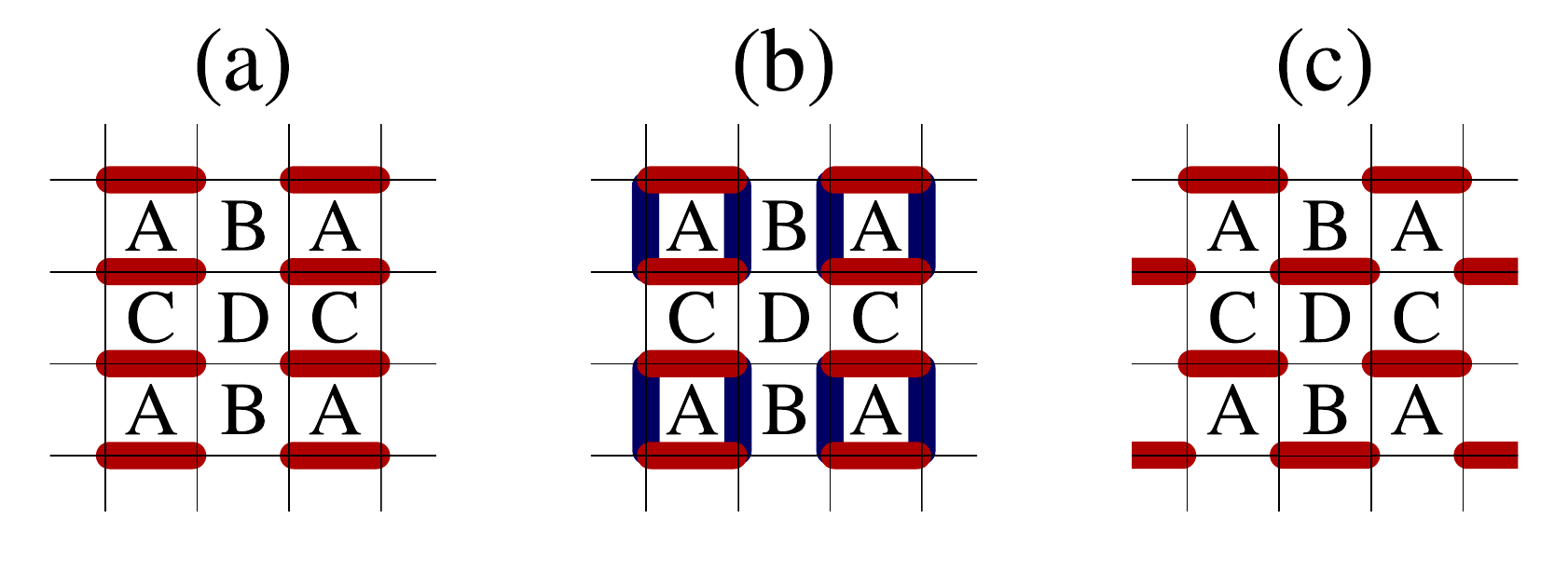}
\end{center}
\caption{Possible ordered phases of the quantum dimer model on the square lattice: (a) columnar phase; (b) plaquette or mixed phase, depending on the relative amplitude between the horizontal and vertical orientations; (c) staggered.}
\label{Dimer_phases}
\end{figure}

Possible ground states of the square-lattice QDM can be divided into two kinds: dimer liquids, which are topologically ordered phases that break no symmetries \cite{Balents2010}, and conventional ordered phases (``dimer solids''), which break lattice symmetries. The ordered states, illustrated in \reffig{Dimer_phases}, can be further divided according to their value of the flux, which
vanishes in the columnar, \reffig{Dimer_phases}(a), plaquette and mixed phases \reffig{Dimer_phases}(b), and is maximized in the staggered phase \reffig{Dimer_phases}(c).

A liquid phase of the dimer model is one that preserves the full symmetry of the lattice, and is characterized by topological order \cite{Balents2010} and fractionalized monomer excitations \cite{Chalker}. According to a result of Polyakov \cite{Polyakov1977}, a quantum \(\mathrm{U}(1)\) liquid phase, i.e., a phase whose long-wavelength description is a \(\mathrm{U}(1)\) gauge theory, is not possible in 2D. It can be shown, however, that the ground state at the RK point, \(v/\t=1\) (and \(v,\t>0\)), can be found exactly, and is a \(\mathrm{U}(1)\) quantum liquid \cite{Fradkin2004}. (This fine-tuned liquid, existing only an isolated point in the phase structure, is consistent with Polyakov's argument.)

To see this, start from an arbitrarily chosen dimer configuration \(c\) and construct the set \(\goC_c\) of all configurations that can be reached from \(c\) by successive plaquette flips. By writing \(\H_{\t,\t}\) as a sum of projectors \cite{Chalker},
\beq
\H_{\t,\t} = \t \sum (\ket{\vertdimer} - \ket{\horidimer} )( \bra{\vertdimer} - \bra{\horidimer})\punc,
\eeq
it is straightforward to show that the equal-amplitude superposition \(\sum_{c' \in \goC_c} \ket{c'}\) is an eigenstate of \(\H_{\t,\t}\), with eigenvalue zero, and so, by the Perron--Frobenius theorem, a ground state. Because plaquette flips cannot change the flux, there is such a zero-energy ground state in each flux sector.

We will mostly be concerned with the RK state at zero flux, constructed by choosing for \(c\) any dimer configuration in this flux sector. This state clearly preserves all symmetries of the Hamiltonian, and is an example of a quantum dimer liquid.

The staggered states are given by those configurations of the dimer model that have no flippable plaquettes. They are therefore trivially eigenstates of the Hamiltonian with zero energy (for any \(v/\t\)), and can be shown to be ground states for all \(v/\t>1\) \cite{Chalker}. Such states saturate the bound on the flux, with \(\Phiv = \{\frac{1}{2}L^2,0\}\) in the configuration shown in \reffig{Dimer_phases}(c), and others related by symmetry. These states break rotation and translation symmetries of the lattice.

For \(v/\t < 1\), the energy is instead minimized by a nontrivial superposition of dimer configurations, and so the ground state depends on \(v/\t\). One therefore expects a discontinuity in the first derivative of the ground-state energy at \(v/\t=1\), and hence a first-order quantum phase transition, according to the standard thermodynamic classification. (This will indeed by confirmed by our results, presented in \refsec{SecResults}, although the presence of the RK point leads to more subtle behavior on the side with \(v/\t < 1\).) Analytical arguments do not determine the ground state for \(v/\t < 1\), however, and numerical approaches must instead be used. The candidates are those with zero flux, illustrated in \reffig{Dimer_phases}(a) and (b).

In the limit \(v \rightarrow -\infty\), the ground state maximizes the number of flippable plaquettes \(N\sub{f}\). The maximal value \(N\sub{f} = \frac{1}{2}L^2\) is achieved by the four configurations with dimers arranged in columns; one is shown in \reffig{Dimer_phases}(a), and the others are related by symmetry. For large negative \(v/t\), the ground state is therefore expected to be an ordered state continuously connected to this limit (and so with the same symmetry), referred to as the columnar phase.

In the same way, the plaquette and mixed phases are continuously connected to product states
\beq[EqProductState]
\ket{\theta} = \prod_{p\in \goP} \left(\ket{\horidimer} \cos\theta + \ket{\vertdimer}\sin\theta\right)\punc,
\eeq
where the product is over a set \(\goP\) of plaquettes tiling the lattice without overlapping, such as those labeled A in \reffig{Dimer_phases}. The plaquette state has \(\theta = \frac{\pi}{4}\), while the mixed state continuously interpolates between the plaquette and columnar (\(\theta = 0\)) states. The plaquette and mixed states include resonances that reduce the kinetic energy, at the expense of decreasing the number of flippable plaquettes \cite{FootnoteProductState} and hence increasing the potential energy. They may therefore compete with the columnar state for \(v/\t\) of order unity.

An order parameter for these phases is provided by the {\it magnetization} \(\Nv\) \cite{Sachdev1989,FootnoteOrderParameter},
\beq[EqMagnetizationDefinition]
N_\mu = \sum_{\rv} (-1)^{r_\mu} d_\mu(\rv)\punc,
\eeq
which points along one of the square axes, \(\pm \deltav_x\) or \(\pm \deltav_y\), in the columnar phase and along one of the four diagonals, \(\pm \deltav_x \pm \deltav_y\), in the plaquette phase, while interpolating between the two in the mixed phase. It vanishes by symmetry at the RK point and also in the staggered states. Note that the magnetization is distinct from the order parameter \(M_X\) used by Banerjee et al.\ \cite{Banerjee}, being naturally defined in terms of the dimers, rather than through height fields. The magnetization \(\Nv\) is well established as an order parameter in the context of classical dimer models \cite{Alet2006}.

The phase diagram for \(v/\t < 1\) has been extensively studied using exact diagonalization (ED) \cite{Sachdev1989,Leung1996} and quantum Monte Carlo (QMC) \cite{Syljuasen2006,Ralko2008,Banerjee}, leading to a variety of contradictory conclusions. Sachdev performed exact diagonalization on lattices up to \(6 \times 6\) and found that the columnar phase extends from \(v/\t = -\infty\) all the way to the RK point (\(v/\t = 1\)). Leung et al.\ extended the calculations to \(8 \times 8\) and concluded that there is an intermediate phase consistent with plaquette order. The same conclusion, though with a different critical value for \(v/\t\), was reached by Sylju{\aa}sen \cite{Syljuasen2006} using projector QMC methods. Ralko et al.\ \cite{Ralko2008}, combining QMC and ED, agreed with the presence of an intermediate phase, but argued that it showed mixed order. Finally, Banerjee et al.\ \cite{Banerjee}, who used a height representation to access larger system sizes than previous MC studies, concluded that there is no intermediate phase, with the columnar phase extending as far as the RK point.

\subsection{Classical dimer model}
\label{SecCDM}

The stochastic CDM that we consider is one with continuous-time Markov dynamics 
within the same set of close-packed dimer configurations. The master equation for the evolution of the probability over configurations can be written in general as
\beq[ME]
\partial_\tt \ket{P_\tt} = \W \ket{P_\tt} ,
\eeq
where $\ket{P_\tt}$ is the probability vector, 
\beq[Pt]
\ket{P_\tt} = \sum_c P_\tt(c) \ket{c} ,
\eeq
with $\left\{ \ket{c} \right\}$ the configuration basis, and $P_\tt(c)$ the probability of configuration $c$ at time $\tt$. The general form of the generator (or master operator) is
\beq[W]
\W = \sum_{c,c'\neq c} w_{c \rightarrow c'} \ket{c'} \bra{c} - \sum_{c} R_c \ket{c} \bra{c} .
\eeq
The positive terms are off-diagonal and encode the possible transitions ${c \rightarrow c'}$ and their rates $w_{c \rightarrow c'}$. The negative terms are diagonal, with $R_c$ the escape rate from configuration $c$, $R_c = \sum_{c'\neq c} w_{c \rightarrow c'}$. The form \eqref{W} guarantees probability conservation: the largest eigenvalue of $\W$ is zero and its left eigenvector is the {\em uniform} (or ``flat'') state: 
\begin{align}
\label{pcons}
\bra{-} \W &= 0\punc,
&
\bra{-} &= \sum_c \bra{c}\punc.
\end{align}

For the specific case of the CDM, a given plaquette, when flippable, flips according to a Poisson process with rate constant \(\gamma\) \cite{Oakes2016}. 
The generator for the dynamics then reads
\beq[MOp]
\W = \gamma \sum(\ket{\vertdimer} \bra{\horidimer} + \ket{\horidimer} \bra{\vertdimer}) - 
\gamma
\sum(\ket{\vertdimer} \bra{\vertdimer} + \ket{\horidimer} \bra{\horidimer})\punc .
\eeq
The terms in the first summation correspond to transitions due to plaquette flips. All allowed transitions have the same rate $\gamma$. The second sum is over the escape rates and ensures conservation of probability. The generator \eqref{MOp} is Hermitian (and hence bistochastic), which means that the uniform state is also the right eigenstate of the zero eigenvalue -- and thus the stationary state of the dynamics, 
\begin{align}
\label{ss}
\W \ket{\rm ss} &= 0 \punc, & 
\ket{\rm ss} &= {\cal N}^{-1} \ket{-} \punc,
\end{align}
where the normalisation is given by $\braket{-|{\rm ss}} = 1 \Rightarrow 
{\cal N} = \braket{-|-}$. As in the case of the Hamiltonian \eqref{QDMH}, the classical dynamics generated by $\W$ conserves the flux for periodic boundary conditions, and thus each flux sector is an irreducible partition of the dynamics. 

At the RK point, $v/\t=1$, the classical generator \eqref{MOp} and the Hamiltonian \eqref{QDMH} become identical up to a sign and an overall factor that is determined by the transition rate \cite{RK1988}, 
\beq[RK]
\W  = -\H_{\gamma,\gamma} .
\eeq
It follows that the ground state of the QDM at the RK point coincides (up to normalisation) with the stationary-state probability of the CDM, i.e., an equal superposition of all dimer configurations: 
\beq[gs]
\ket{\gs}_{\rm RK} = {\cal N}^{-1/2} \ket{-} .
\eeq 
This shows that it is possible to probe the ground state properties of quantum dimers at the RK point from the stationary state dynamics of classical dimers.

\subsection{Trajectory ensembles and dynamical large deviations}

The dynamics generated by \eqref{MOp} is realised in terms of stochastic trajectories. For a continuous time Markov chain such as we are considering, a trajectory $\omega_\tt$ of overall time extension $\tt$ is a sequence of configurations and jumps between them, 
\beq[traj]
\omega_\tt = (c_0, c_{\tt_1}, \ldots, c_{\tt_K})\punc,
\qquad
0 < \tt_1 < \cdots < \tt_K < \tt\punc,
\eeq
where $\tt_i$ ($i=1,\ldots,K$) indicate the times at which jumps between configurations occur. Between jumps the configuration remains unchanged, so that from the time of the last jump, $\tt_K$, to the final time, $\tt$, the configuration in trajectory \eqref{traj} would be $c_{\tt_K}$.

The dynamics generates an {\em ensemble of trajectories}, defined as the set of all possible trajectories \eqref{traj} and their probabilities to occur, $\pi(\omega_\tt)$. The ensemble of trajectories contains the information about all possible time correlations, and thus encodes more information about the dynamics than the probability \(P(c,\tt')\). In particular, the latter is obtained from $\pi$ by summing over all stochastic trajectories (i.e., by {\em contraction}),
\beq[cont]
P(c,\tt') = \sum_{\omega_\tt} \pi(\omega_\tt) \delta\pmb{(}c_{\tt'}(\omega_\tt) - c\pmb{)} \punc,
\eeq
where \(c_{\tt'}(\omega_\tt)\) is the configuration at time \(\tt'\) in trajectory \(\omega_\tt\).

The properties of trajectories can be catalogued by trajectory observables. The simplest of these is the {\em dynamical activity} 
\cite{DeformationPaper1,DeformationPaper2,DeformationPaper3,Baiesi2009} defined as the number of configuration changes in a trajectory, which we will denote by the symbol $\hat{K}$ when acting on trajectories. For example, for the trajectory in \refeq{traj} we have $\hat{K}(\omega_\tt) = K$, as there are a total of $K$ jumps in that trajectory.

The activity is a trajectory order parameter -- it is extensive in both system size and observation time -- and is the natural quantifier of the overall ``amount of motion''. It does this quantification of motion in a ``structure-agnostic'' way; that is, it does not assume any particular configurational property underlying fast or slow relaxation. In the course of a trajectory the activity simply increases by one unit every time the system changes its configuration, irrespective of the nature of those changes. It is particularly well suited for glassy systems where there is no obvious structural order parameter associated to glassiness; see e.g.~\refcite{Chandler2010}.

Associated with the ensemble of trajectories is a corresponding distribution for trajectory order parameters such as the activity, which has probability distribution
\beq[PK]
P_\tt(K) = \sum_{\omega_\tt} \pi(\omega_\tt) \delta[\hat{K}(\omega_\tt) - K] \sim \ee^{-\tt \varphi(\frac{K}{\tt})} \punc.
\eeq
The approximate equality is the large deviation (LD) form of the probability that is applicable at long times (as long as the correlation times of the dynamics remain finite) \cite{Touchette2009,DeformationPaper1,DeformationPaper2,DeformationPaper3}. At these long times the statistics of $K$ are determined by the LD {\em rate function} $\varphi(k)$ which can be thought of as an entropy density in the space of trajectories \cite{Touchette2009,DeformationPaper1,DeformationPaper2,DeformationPaper3}. 

Equivalent information to that found in $P_\tt(K)$ is contained in the moment generating function (MGF) \cite{Touchette2009,DeformationPaper1,DeformationPaper2,DeformationPaper3}, 
\beq[Zs]
Z_\tt(s) 
= \sum_K P_\tt(K) \ee^{-s K}
= \sum_{\omega_\tt} \pi(\omega_\tt) \ee^{- s \hat{K}(\omega_\tt)} 
\sim \ee^{\tt \theta(s)} .
\eeq
This function generates the moments of $K$, via
$\braket{K^n} = (-1)^n \partial_s^n Z_\tt(s)|_{s=0}$. Just as for the probability, the MGF has a LD form at long times, given by the approximate equality in \refeq{Zs}. The function $\theta(s)$ is the scaled cumulant generating function (SCGF; its derivatives evaluated at $s = 0$ give the cumulants of $K$ scaled by time), and  plays the role of a free-energy density for trajectories \cite{DeformationPaper1,DeformationPaper2,DeformationPaper3}.

One way to interpret Eqs.\ \eqref{PK} and \eqref{Zs} is in terms of {\em conditioned} and {\em biased} trajectory ensembles \cite{Chetrite2015}. The delta function in \eqref{PK} restricts the sum to trajectories which have total activity $K$. This is analogous to a microcanonical ensemble which restricts configurations to a fixed energy. In turn, in \eqref{Zs} the sum is over all trajectories but the probabilities are exponentially biased (or {\em exponentially tilted}). Here $K$ is only indirectly controlled by $s$. This is analogous to a canonical ensemble of configurations controlled by an inverse temperature. As in the static case for large volume, the two trajectory ensembles are equivalent for long times. In particular, the rate function and the SCGF are related by a Legendre transform \cite{Touchette2009,DeformationPaper1,DeformationPaper2,DeformationPaper3},
\beq[LT]
\theta(s) = - \min_{k} \left[ \varphi(k) + s k \right] .
\eeq

\subsection{$s$-ensemble}

While both trajectory ensembles encode the same information about the dynamics, the ``canonical'' ensemble, \refeq{Zs}, is more practical to study. This is sometimes called the {\em $s$-ensemble} \cite{Hedges2009}. The probabilities of trajectories are exponentially tilted from the natural ones of the dynamics as
\beq[sens]
\pi_s(\omega_\tt) = \frac{\pi(\omega_\tt) \ee^{- s \hat{K}(\omega_\tt)}}{Z_\tt(s)} . 
\eeq
We denote averages of trajectory observables $\hat{A}$ in this ensemble by $\braket{\hat A}_s$, which in terms of the averages of the original dynamical ensemble read
\beq[avs]
\braket{\hat A}_s = \frac{\braket{\hat{A} \ee^{- s \hat{K}}}}{Z_\tt(s)} . 
\eeq
The power of the $s$-ensemble is that it allows a full characterisation of the 
dynamics beyond typical behaviour by tuning $s$ away from $s=0$. In particular, at long times, the analytic structure of the SCGF determines the phase structure of the dynamics. This is analogous to what occurs with the free-energy in static problems. 

One reason that the $s$-ensemble is more tractable is that the dynamical partition sum \eqref{Zs} can be written in ``transfer matrix'' form,
\beq[Zss]
Z_\tt(s) = \bra{-} \ee^{\tt {\mathbb W}_s} \ket{\text{i}} \punc,
\eeq
where the probability vector $\ket{\text{i}}$ represents the distribution from which the initial state is drawn. The operator ${{\mathbb W}_s}$ is a deformation or tilt of the original dynamical generator that reads (for the case of tilting against the activity) \cite{DeformationPaper1,DeformationPaper2,DeformationPaper3}
\beq[Ws]
\W_s = \ee^{-s} \sum_{c,c'\neq c} w_{c \rightarrow c'} \ket{c'} \bra{c} - \sum_{c} R_c \ket{c} \bra{c} \punc.
\eeq

The equivalence between \refeqand{Zs}{Zss} can be proved in the following way. If we write the master operator as $\W = \sum_\mu \J_\mu - \RR$, where $\J_\mu$ denotes the off-diagonal parts of $\W$ that are responsible for all the possible transitions $\mu$, and $\RR$ is the diagonal part of $\W$ with the escape rates, cf.\ \refeq{W}, the probability of $\pi(\omega_\tau)$ of a trajectory such as \refeq{traj} can be written as 
\begin{equation}
\pi(\omega_\tau)
=
\langle c_{\tau_K} |
e^{- (t-t_K) \RR}
\,
\J_{\mu_K} e^{- (t_K-t_{K-1}) \RR} 
\cdots
\J_{\mu_{1}} e^{- t_{1} \RR} | c_0 \rangle 
, \nonumber
\end{equation}
In \refeq{Zs} we have the exponentially tilted probability
\begin{align}
\pi(\omega_\tt) \ee^{- s \hat{K}(\omega_\tt)}
&=
\langle c_{\tau_K} |
e^{- (t-t_K) \RR}
\,
e^{-s}
\J_{\mu_K} 
\nonumber \\
&\times 
e^{- (t_K-t_{K-1}) \RR} 
\cdots
e^{-s}
\J_{\mu_{1}} e^{- t_{1} \RR} | c_0 \rangle 
. \nonumber
\end{align}
Summing over all trajectories to obtain \refeq{Zs} then gives
\begin{align}
Z_\tt(s) &= \sum_{K=0}^\infty 
\,
\sum_{\mu_1} \cdots \sum_{\mu_K} 
\,
\int_0^\tt d\tt_1 
\cdots
\int_{\tt_{K-1}}^\tt d\tt_K
\label{mps} \\
&
\langle - |
e^{- (\tt-\tt_K) \RR}
\,
e^{-s}
\J_{\mu_K} e^{- (\tt_K-\tt_{K-1}) \RR} 
\cdots
e^{-s}
\J_{\mu_{1}} e^{- \tt_{1} \RR} | {\rm i} \rangle 
, \nonumber
\end{align}
which is \refeq{Zss} expressed as a Dyson series for the tilted generator \refeq{Ws}.

In contrast to $\W$, defined in \refeq{W}, the tilted operator \(\W_s\) does not define a probability conserving dynamics for $s \neq 0$. In fact, its largest eigenvalue is $\theta(s)$, and thus the problem of computing the dynamical partition sum reduces to that of maximising \eqref{Ws}. The long-time average activity, obtained from the SCGF as 
\beq[Ks] 
\lim_{\tt \to \infty} \frac{\braket{\hat K}_s}{\tt} = - \theta'(s) \punc.
\eeq
serves as the dynamical order parameter that helps classify the dynamical phase behaviour, with associated susceptibility,
\beq[chi] 
\chi_s = \lim_{\tt \to \infty} \frac{\braket{\hat K^2}_s - \braket{\hat K}_s^2}{\tt} = \theta''(s) \punc.
\eeq

\subsection{Connection to QDM}
\label{SecConnectionToQDM}

For the specific case of the dimer model, the tilted operator reads 
\beq[MOps]
\W_s = \ee^{-s} \gamma \sum(\ket{\vertdimer} \bra{\horidimer} + \ket{\horidimer} \bra{\vertdimer}) 
- \gamma \sum(\ket{\vertdimer} \bra{\vertdimer} + \ket{\horidimer} \bra{\horidimer})\punc .
\eeq
We see from \refeq{QDMH} that this coincides with the general QDM Hamiltonian, if we identify $\t = \ee^{-s} \gamma$ and $v = \gamma$:
\beq[WsH]
\W_s = - \H_{\ee^{-s} \gamma,\gamma}\punc.
\eeq
Changing $s$ is equivalent to changing the ratio $v/t$, and so the properties of the $s$-ensemble of classical trajectories of the CDM are directly related to those of the QDM.

To be specific, consider the ground-state expectation value of a quantum observable represented by an operator \(\dsQ\) that is diagonal in the basis of dimer configurations. To find this, we evaluate \(\dsQ\) in the configuration at the midpoint of each trajectory and average over trajectories with \(s\) weighting. The latter gives \(\langle \dsQ(\frac{\tt}{2}) \rangle_s = \sum_{c} P_s(c,\frac{\tt}{2}) \bra{c}\dsQ\ket{c}\), where \(P_s\) is given by \refeq{cont} with the replacement \(\pi\rightarrow\pi_s\). Using \refeq{sens}, and applying the same steps that led from \refeq{Zs} to \refeq{Zss}, one finds
\beq
\langle \dsQ(\tfrac{\tt}{2}) \rangle_s = \dfrac{\bra{-} \ee^{\frac{\tt}{2} {\mathbb W}_s} \dsQ \ee^{\frac{\tt}{2} {\mathbb W}_s} \ket{\text{i}}}{\bra{-} \ee^{\tt {\mathbb W}_s} \ket{\text{i}}}\punc.
\eeq
The expectation value of \(\dsQ\) in \(\ket{\gs}\), the ground state of \(\H_{\ee^{-s} \gamma,\gamma}\), is therefore given by the limit of large \(\tt\),
\beq
\bra{\gs}\dsQ\ket{\gs} = \lim_{\tt\rightarrow\infty} \langle \dsQ(\tfrac{\tt}{2}) \rangle_s\punc.
\eeq
Note that this is true for arbitrary initial distribution \(\ket{\text{i}}\), as long as the overlap \(\braket{\gs\vert \text{i}}\) is nonzero.

In what follows we exploit the relationship between the quantum and classical models, and connect the ground state phase diagram of the QDM to the properties of CDM trajectories explored numerically via path-sampling methods.

\section{Trajectory Sampling of classical dimers}

\subsection{Transition path sampling} 

The main difficulty in sampling $s$-ensemble trajectories is the usual one associated with calculating exponential averages \cite{Gingrich2015}: the trajectories that are easy to generate with the normal dynamics at \(s=0\), \refeq{MOp}, are not the relevant ones for the biased ensemble at $s \neq 0$, \refeq{sens}, and the latter are exponentially rare in the original dynamics. Since the tilted operator \eqref{Ws} is not a dynamical generator, there is no simple way to generate the relevant trajectories directly. 

As in a static context (think for example of sampling the equilibrium of a spin system at finite temperature), this is resolved by importance sampling \cite{Chandler1987}. In the case of trajectories one such importance sampling scheme is Transition Path Sampling (TPS) \cite{TPSreview}. TPS is a set of numerical techniques developed for generating rare trajectories that are infrequent enough that their observation through brute force simulations is unfeasible. Using original dynamics to generate trial trajectories, TPS performs a biased random walk through trajectory space towards the region of rare trajectories that exhibit the desired behavior. TPS is particularly appropriate for sampling dynamics with detailed balance, as in the case of the CDM, \refeq{MOp}.

The basic idea behind TPS is similar to that of Markov chain Monte Carlo but applied to trajectories \cite{TPSreview}. In its simplest form the procedure is as follows: (i) Given a trial trajectory, a new trajectory is proposed (as we describe below in detail). (ii) The proposed trajectory is accepted or rejected according to a Metropolis criterion. Since in our case we want to sample \refeq{sens}, the key quantity is the change in overall activity $\Delta K$ between the old trajectory and the new one. As in standard Metropolis, if $\Delta K < 0$ the new trajectory is always accepted; if $\Delta K > 0$, it is only accepted with probability $\ee^{-s \Delta K}$. The procedure is repeated until the ensemble of trajectories thus generated converges to the \(s\)-ensemble.

\begin{figure*}
\begin{center}
\includegraphics[width=\textwidth]{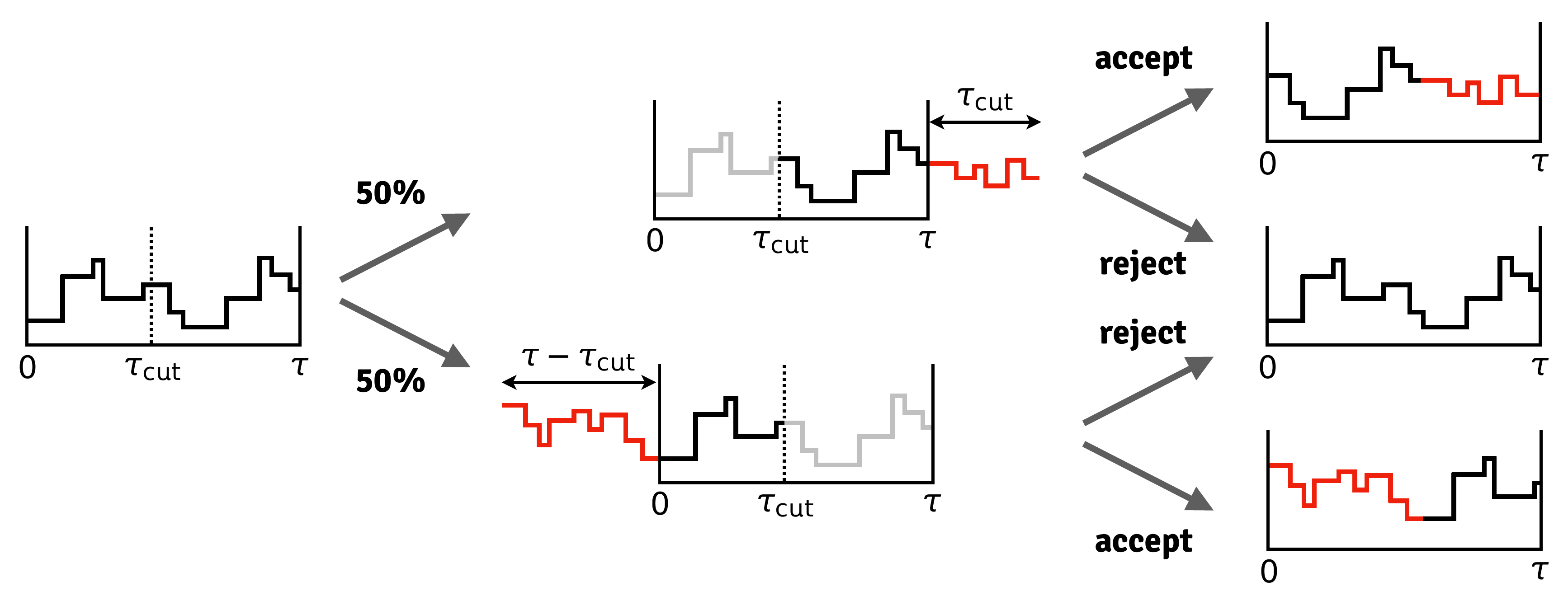}
\end{center}
\caption{
An illustration of the TPS shifting method. The current trajectory, of total time extension $\tt$, is shown on the left. A cut time $\tt_{\rm cut}$ is chosen randomly and uniformly between initial time $0$ and final time $\tt$. With equal probabilities, a new trajectory is proposed via the {\em shift backwards} (centre, top) or via the {\em shift forwards} (centre, bottom) procedures. For a shift backwards (centre, top), the portion of the original trajectory {\em after} $\tt_{\rm cut}$ is kept (black) while the portion before is discarded (grey). From the end of the retained trajectory segment a new segment of extent $\tt_{\rm cut}$ is generated (red) with the usual dynamics. The proposed new trajectory is formed of the old black segment and the new red segment (all shifted in time backwards by $\tt_{\rm cut}$). 
For a shift forwards (centre, bottom), the portion of the original trajectory {\em before} $\tt_{\rm cut}$ is kept (black) while the portion after is discarded (grey). Starting from the initial condition of the retained segment a new segment of extent $\tt-\tt_{\rm cut}$ is generated (red) with the usual dynamics and time reversed (which is possible -- and efficient -- in our case due to detailed balance in the CDM dynamics). The proposed new trajectory is formed of the new red segment and the old black segment (all shifted in time forwards by $\tt-\tt_{\rm cut}$). The proposed new trajectories are then accepted or rejected according to a Metropolis criterion as described in the main text, as sketched on the right.}
\label{TPS cartoon}
\end{figure*}

The non-trivial step is (i). Various methods for generating trajectories have been proposed \cite{TPSreview} that guarantee ergodicity in trajectory space and are reasonably efficient. In particular, we use the {\em shifting} method \cite{TPSreview}, described in \reffig{TPS cartoon}: Given an initial trajectory, a new trajectory is proposed by choosing an arbitrary cut time, $\tt_{\rm cut}$ (chosen uniformly between 0 and $\tt$), keeping either the portion of the trajectory {\em after} the cut, $\tt > \tt_{\rm cut}$, and shifting {\em back} to time $0$; or keeping the portion of the trajectory {\em before} the cut, $\tt < \tt_{\rm cut}$, and shifting {\em  forward} to $\tt$. These two options are chosen with equal probability.

The remaining part of the old trajectory is discarded and has to be replaced by a new partial trajectory. In the case of a backward shift, the missing part is that from $\tau_{\rm cut}$ onwards; the new portion is obtained by shooting a new trajectory with the original dynamics from the configuration at $\tt_{\rm cut}$ (after the shift, i.e., the final configuration of the original trajectory) up to the final time $\tt$.  In the case of a forward shift, the missing part is between time $0$ and $\tt_{\rm cut}$; to fill it one shoots a new trajectory with the original dynamics {\em forwards} starting from the configuration at $\tt_{\rm cut}$ (after the shift, i.e., the initial configuration of the original trajectory) for a length $\tt_{\rm cut}$ and then inverts time. This is a valid procedure in the case of detailed balance dynamics \cite{TPSreview}.

For step (ii) we need the difference in activities between the trajectories. The current and proposed trajectory share the portion that has been shifted, and so the difference in activity comes only from the newly generated part. Since $\Delta K$ is a time-extensive quantity, acceptance will be suppressed exponentially. The fundamental limitation of this version of TPS for sampling long trajectories is that ergodicity in the dynamics implies that proposed trajectories diverge exponentially fast from their seed. This should be contrasted with sampling a spin model, for example, where new configurations can be proposed by flipping just a single spin, thus preventing the energy difference from growing with system size.

The above means that, while simple TPS can sample the $s$-ensemble much more efficiently than brute force sampling (and indeed has been used successfully in this context before \cite{Hedges2009,Elmatad2010,Speck2012}), the exponential cost of sampling long times may render it impractical. An alternative to TPS is the {\em cloning method} \cite{Giardina2011} adapted from quantum diffusion Monte Carlo, which however also suffers from a similar exponential cost \cite{Ray2017}. Below we discuss how to parially overcome this limitation for TPS by exploiting umbrella sampling techniques to the trajectory context \cite{Nemoto2016,Klymko2017,Ray2017,Ray2017b}.

\subsection{TPS and trajectory umbrella sampling}
\label{SecTPSUmbrella}

The idea of umbrella sampling is the following \cite{Nemoto2016,Klymko2017,Ray2017,Ray2017b}. We wish to compute exponential averages of the form
\beq[expav]
\braket{\ee^{- s \hat{K}}} = \sum_{\omega_\tt} \pi(\omega_\tt) \ee^{- s \hat{K}(\omega_\tt)} \punc,
\eeq
where $\pi(\omega_\tt)$ is the probability at which trajectories are generated by the original dynamics. Consider now an alternative (and proper stochastic) {\em reference} dynamics where the same trajectories are generated with probability $\pi_{\rm ref}(\omega_\tt)$. We can rewrite \refeq{expav} as
\beq[expav2]
\braket{\ee^{- s \hat{K}}} = \sum_{\omega_\tt} \pi_{\rm ref}(\omega_\tt) 
\frac{\pi(\omega_\tt)}{\pi\sub{ref}(\omega_\tt)}
\ee^{- s \hat{K}(\omega_\tt)} 
= 
\braket{\R \ee^{- s \hat{K}}}\sub{ref}\punc,
\eeq
where 
\beq[R]
\R(\omega_\tt) = \frac{\pi(\omega_\tt)}{\pi\sub{ref}(\omega_\tt)}\punc,
\eeq
This simply means that the average of the trajectory observable $\ee^{- s \hat{K}}$ over the original dynamics is the same as the average of the trajectory observable $\R \ee^{- s \hat{K}}$ over the reference dynamics, \(\braket{\cdots}\sub{ref}\). The ``umbrella'' $\R$ compensates for the change of probability. 

This can be useful in the following way. Given a reference dynamics, we would estimate \eqref{expav2} by an empirical average over $N_{\rm sp}$ sample trajectories,
\beq[sp]
\braket{\ee^{- s \hat{K}}} = \braket{\R \ee^{- s \hat{K}}}_{\rm ref}
\approx \frac{1}{N_{\rm sp}} \sum_{\alpha=1}^{N_{\rm sp}} \R(\omega^\alpha) \ee^{- s \hat{K}(\omega^\alpha)} .
\eeq
The sampling error is given by the variance of the average squared of the empirical average,
\begin{align}
\varepsilon^2 
&= 
\frac
{
{\rm var}_{\rm ref}\left(  
\frac{1}{N_{\rm sp}} \sum_{\alpha=1}^{N_{\rm sp}} \R(\omega^\alpha) \ee^{- s \hat{K}(\omega^\alpha)}
\right)
}
{
\Braket 
{
\frac{1}{N_{\rm sp}} \sum_{\alpha=1}^{N_{\rm sp}} \R(\omega^\alpha) \ee^{- s \hat{K}(\omega^\alpha)}
}^2_{\rm ref}
}
\nonumber
\\
&=
\frac{1}{N_{\rm sp}}
\left(
\frac
{
\braket{\R^2 \ee^{-2 s \hat{K}}}_{\rm ref}
}
{
\braket{\R \ee^{- s \hat{K}}}_{\rm ref}^2
}
-1
\right)
\nonumber
\\
&=
\frac{1}{N_{\rm sp}}
\left(
\frac
{
\braket{\R \ee^{-2 s \hat{K}}}
}
{
\braket{\ee^{- s \hat{K}}}^2
}
-1
\right) ,
\label{e2}
\end{align}
where we have used \refeq{expav2} between the second and third lines to recast $\varepsilon^2$ in terms of the original averages. 

Consider the case where the reference dynamics is just the original one, $\R=1$. The sampling error reads,
\beq[e22]
\varepsilon^2 
=
\frac{1}{N_{\rm sp}}
\left(
\frac
{
\braket{\ee^{-2 s \hat{K}}}
}
{
\braket{\ee^{- s \hat{K}}}^2
}
-1
\right) 
\approx 
\frac{\ee^{\tt [\theta(2s)-2\theta(s)]}}{N_{\rm sp}} , 
\eeq
where in the last line we have used \refeq{Zs} for long times. The convexity of the SCGF function implies that $\theta(2s) \geq 2\theta(s)$ always, and the error diverges exponentially with time. This is why sampling with the original dynamics is inefficient, and accurate estimation requires exponentially many trajectories $N_{\rm sp}$. The aim is therefore to find an alternative reference dynamics which makes the convergence of \eqref{sp} more efficient.

\subsubsection{Ideal reference dynamics: generalised Doob transformation}

The reweighting factor \eqref{R} for a trajectory such as \eqref{traj} reads,
\beq[RR]
\R(\omega_\tt) = 
\ee^{-\tt_1 \Delta R_{c_0}} 
\frac{w_{c_0 \to c_{\tt_1}}}{w_{c_0 \to c_{\tt_1}}^{\rm ref}} 
\ee^{-(\tt_2-\tt_1) \Delta R_{c_{\tt_1}}} 
\cdots ,
\eeq
where $\Delta R_c = R_c - R_c^{\rm ref}$. It contains exponential factors for all the time periods between jumps and ratios of the transition probabilities for each jump in the trajectory. 
The ideal choice for a reference dynamics would be one that cancels the exponential growth of the numerator in the first term of \eqref{e2}. This is known to be given by the {\em generalised Doob transform} \cite{Jack2010,Chetrite2015,Garrahan2016}, which maps the tilted generator \eqref{Ws} to a new stochastic operator $\tilde{\W}$ whose natural trajectories are those of the $s$-ensemble. For long observation times the transformation is obtained in the following way. 

From the components $l_c$ of the left eigenstate of $\W_s$, 
\beq[ls]
\bra{l_s} \W_s = \bra{l_s} \theta(s) ,
\eeq
we construct a diagonal matrix $L_s$, such that $\bra{-} L_s = \bra{l_s}$.  
We then define 
\beq[Doob]
\tilde{\W} = L_s \left(\W_s - \theta_s \mathbb{I} \right) L_s^{-1} \punc,
\eeq
where \(\mathbb{I}\) is the identity operator. \(\W_s\) is stochastic, 
\beq[Doob2]
\bra{-} \tilde{\W} = 0 ,
\eeq
and for long times is guaranteed to generate the same trajectories as those of the $s$-ensemble. 

The transition rates in $\tilde{\W}$ are given by 
\beq[WD]
\tilde{w}_{c\to c'} = \left( \frac{l_{c'}}{l_{c}} \right) \ee^{-s} {w}_{c\to c'} ,
\eeq
while the escape rates coincide, up to a shift, with the original ones,
\beq[RD]
\tilde{R}_{c} = R_c - \theta(s) .
\eeq
If the reference dynamics is the one generated by $\tilde{\W}$ the reweighing factor then reads,
\beq[RRD]
\R = \ee^{\tau \theta(s)} \ee^{s \hat{K}} \frac{l_{c_0}}{l_{c_\tt}} .
\eeq
We see that this form of $\R$ cancels the exponential averaging in the numerator of \eqref{e2} and the error is no longer exponential in time. 
This means that an $s$-tilted expectation value like \eqref{avs} can be computed by simply running the dynamics with $\tilde{\W}$ \cite{Nemoto2016,Klymko2017,Ray2017,Ray2017b}.

\subsubsection{Effective reference dynamics}
\label{SecEffectiveReferenceDynamics}

While the ideal reference dynamics is provided by the Doob transformed generator $\tilde{\W}$, this is not a useful solution in practice, as one needs to diagonalise $\W_s$ first, which amounts to solving the problem exactly. 
Nevertheless, the form of the ideal transition rates \eqref{WD} can help guide the definition of convenient approximations for the reference dynamics that are practical \cite{Nemoto2016,Klymko2017,Ray2017,Ray2017b}.

We will consider the transition rates for the reference dynamics that have the form of \eqref{WD} 
\beq[Wref]
w_{c\to c'}^{\rm ref} = \left( \frac{\ell_{c'}}{\ell_{c}} \right) 
\ee^{-s} {w}_{c \to c'} .
\eeq
The aim is to find a vector $\bra{\ell}$ that approximates the exact $\bra{l_s}$ and is also tractable numerically. The associated escape rates, 
\beq[Rref]
R_{c}^{\rm ref} = \ee^{-s} \sum_{c' \neq c} \left( \frac{\ell_{c'}}{\ell_{c}} \right) {w}_{c \to c'} , 
\eeq
in general will not be a uniform shift from the original ones as in \eqref{RD}.
The reweighing factor is (up to boundary terms)
\beq[gref]
\R = e^{s \hat{K}} e^{ - \int_0^\tt d\tt' \Delta R_{c(\tt')}} .
\eeq
which together with \eqref{e2} indicates that, in contrast to the Doob-transformed dynamics, in general sampling will be exponentially difficult. Despite this, a judicious choice of $\bra{\ell}$ that reasonably approximates $\bra{l_s}$ improves sampling significantly, as we now show.

The exact $l_c$ \eqref{ls} are functions of the whole configuration $c$ and in general may have correlations at large distances. A simple approximation is to assume a short-range correlated form for $\ell_c$ and write them as a products of local factors. These local factors could in turn be of the exact form for small-enough local regions. We pursue this approach by considering $2\times2$ neighbourhoods with open boundary conditions, as described in \refapp{app_open_2by2}. 
This leads to $\ell_c$ depending on the configuration only through the total number $N_{\rm f}(c)$ of flippable plaquettes,
\beq[ellD]
\ell_c = e^{D N_{\rm f}(c)}, 
\eeq
where the constant $D$ parametrises the function.

Putting this all together, the sampling proceeds as follows: The reference dynamics we use is given by the transition rates 
\beq[WrefD]
w_{c\to c'}^{\rm ref} = e^{D [N_{\rm f}(c')-N_{\rm f}(c)]} 
e^{-s} {w}_{c \to c'} , 
\eeq
and escape rates 
\beq[RrefD]
R_{c}^{\rm ref} = e^{-s} \sum_{c' \neq c} e^{D [N_{\rm f}(c')-N_{\rm f}(c')]} 
{w}_{c \to c'} . 
\eeq
In order to sample the original dynamics tilted by $e^{-s \hat{K}}$ we have to sample the reference dynamics tilted by $\R e^{-s \hat{K}}$, see \eqref{expav2}, which we write as
\beq[g]
\hat{g} = e^{-D [N_{\rm f}(c_{\tt})-N_{\rm f}(c_0)]- \int_0^\tt d\tt' \left( R_{c(\tt')} - R_{c(\tt')}^{\rm ref} \right)}.
\eeq
In order to account for this tilting we use TPS with the reference dynamics, with an acceptance rate for trajectories given by 
\beq[min]
\Gamma_{\rm acc}(\omega \to \omega') = \min\left(1,\frac{\hat{g}(\omega')}{\hat{g}(\omega)}\right).
\eeq

\subsubsection{Optimization of reference dynamics}
\label{SecOptimizationOfReferenceDynamics}

The reference dynamics above is parametrised by the constant $D$. While the dynamics is based on the Doob transformation for the open $2 \times 2$ problem, there is no reason why the value of $D$ that corresponds to the exact solution for the small system will provide the optimal dynamics for the large system. The reference dynamics can be optimised by choosing the value of $D$ that maximises the trajectory acceptance rate $\Gamma_{\rm acc}$ in the TPS simulations. 

\begin{figure}
\begin{center}
\includegraphics[width=.8\columnwidth]{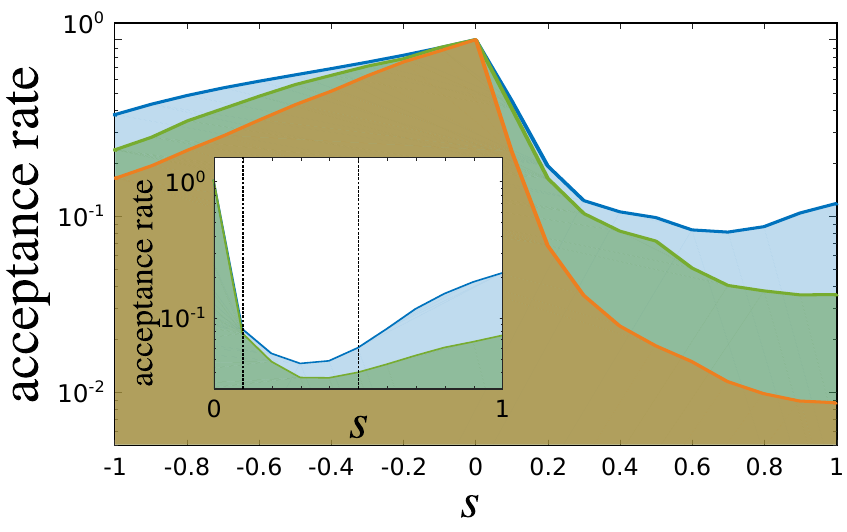}
\end{center}
\caption{
Comparison of TPS acceptance rates for $L=6$ as a function of $s$. The orange curve corresponds to the TPS acceptance when using the original dynamics. The green curve is for the reference dynamics with $D=D_s$ from the \twobytwo\ approximation of \refapp{app_open_2by2}. The blue curve corresponds to the optimal value of $D$ found from exploring the acceptance rate landscape. The data shown is for trajectories of length $\tt = 50$ for $s > 0$, and $\tt = 5$ for $s \le 0$ (convergence in time is much faster on the active side $s\le 0$). Each point shown corresponds to  $5\times10^6$ attempted TPS moves. The optimised $D$ values used are $D = 0.25\,s$ for $s<0$, and $D = s$ for $s>0$. Inset: Acceptance rates for $L=12$ in the region where the active--inactive transition occurs for this system size (see \reffig{FigActivity}).
}
\label{racc_vs_S}
\end{figure}
Finding the optimal value of $D$ is a case of exploring the landscape of acceptance rates $\Gamma\sub{acc}$. Figure \ref{racc_vs_S} compares the acceptance rate as a function of $s$ when using the original dynamics to that obtained when using the effective reference dynamics, Eqs.\ (\ref{ellD}--\ref{RrefD}) with $D_s$ obtained as in \refapp{app_open_2by2}, and with a value of $D$ that optimises even further the acceptance. This latter optimal value of $D$ is obtained for each $s$ by starting from $D=D_s$ and progressively changing $D$ until a maximum of $\Gamma_{\rm acc}$ is reached. The effective reference dynamics defined by this optimal $D$ is then the one used for obtaining the corresponding $s$-ensemble.

\section{Results}
\label{SecResults}

\begin{figure*}
\includegraphics[width=\textwidth]{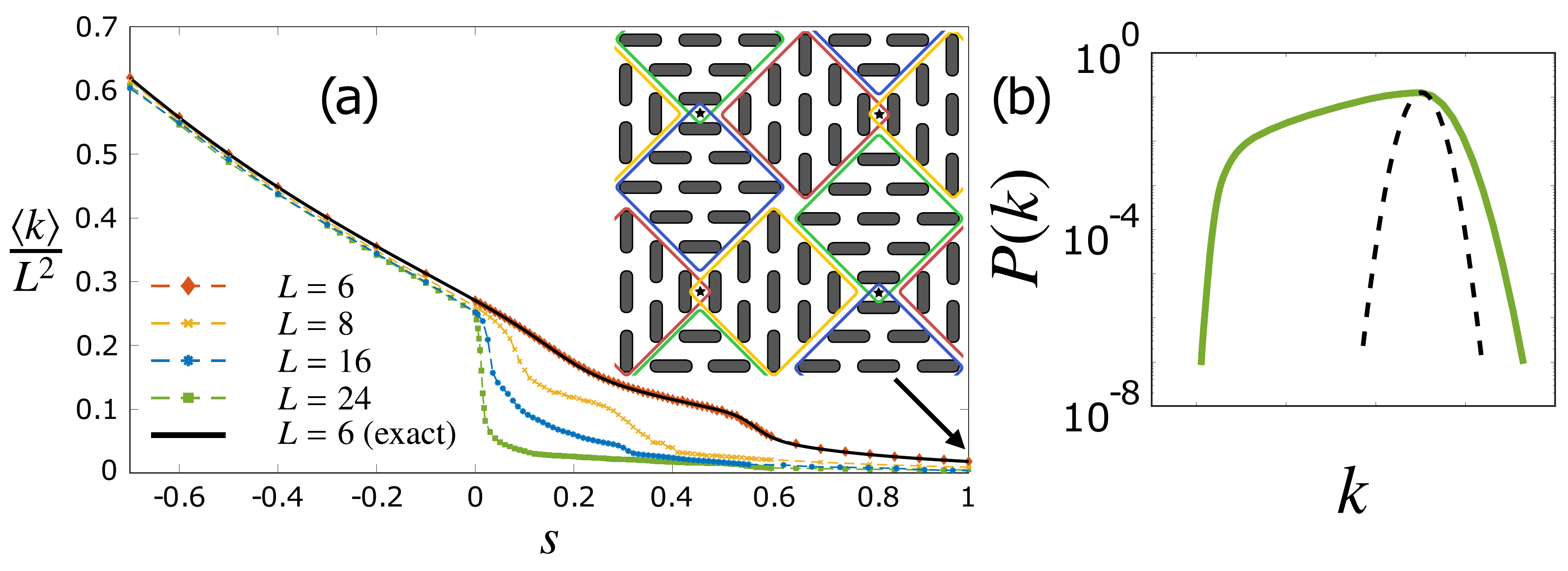}
\caption{(a) Activity rate \(\langle k \rangle = \langle K \rangle / \tt\) as a function of \(s\), for various system sizes \(L\). Symbols show MC results, while the solid black line is from exact diagonalization for \(L=6\). The activity converges quickly for \(s \le 0\), but is strongly system-size dependent for \(s>0\). For larger sizes, there is an increasingly sharp drop in \(\langle k\rangle\) at \(s=0\), suggesting a first-order transition in the thermodynamic limit. Inset: Example (for an \(8\times 8\) lattice) of the minimally flippable zero-flux configurations, which dominate the dynamics at large positive \(s\). The colored squares show domains within which the dimers (stadium shapes) form a staggered arrangement with maximal local contribution to the flux \(\Phiv\). The eight domains, two of each orientation, have equal size, and so the total flux of the configuration is zero. There are four flippable plaquettes, indicated with stars (\(\star\)), the minimum possible number for a zero-flux configuration \cite{OakesThesis}. (b) Distribution of activity rate for \(L = 24\) and \(s=0\) (solid green curve) compared with a Gaussian distribution of the same curvature at the maximum (dashed black). The broadening, particularly on the low-activity side, is a reflection of the sudden drop in \(\langle k\rangle\) at \(s = 0^+\).}
\label{FigActivity}
\end{figure*}

Figure~\ref{FigActivity} shows the mean activity rate \(\langle k\rangle = \langle K \rangle/\tt\) evaluated across a range of \(s\) values, with the results for the smallest system size, \(L=6\), compared with exact diagonalization. The agreement between these results confirms that the method has converged, at least for this \(L\), and demonstrates that it is able to resolve detailed features of the dynamics for \(s>0\). For larger system sizes, there is an increasingly sharp drop in the activity at \(s=0\). The activity is related, by \refeq{Ks}, to the first derivative of the thermodynamic potential \(\theta(s)\) (or equivalently, the quantum ground-state energy), and so this indicates a first-order transition at \(s=0\), as expected from the analytical arguments in \refsec{SecQDM}. (Note that our simulations are restricted to the zero-flux sector, but this is not expected to change the thermodynamic properties in the limit \(L\rightarrow\infty\).) The activity histogram, shown in the inset, shows the characteristic broadening, compared with a Gaussian distribution, expected for such a transition.

As discussed in \refsec{SecQDM}, the ground state for positive \(s\) (\(v/\t>1\)) is fully staggered and so has maximal flux. Our simulations are, however, restricted to the zero-flux sector, and so the long-time dynamics is dictated by the ground state within this sector. For \(s>0\), the activity decreases in a sequence of rounded steps, which are strongly system-size dependent. This is apparently a consequence of commensuration effects, as different low-flippability configurations are favored depending on the precise value of \(s\). For large positive \(s\), the system is mostly restricted to the minimally flippable zero-flux configurations, which, as illustrated in \reffig{FigActivity}(a), have precisely \(4\) flippable plaquettes for any system size \cite{OakesThesis}. Second-order perturbation theory gives the ground-state energy as \(-\theta(s) = -2\ee^{-2s}\), which leads, using \refeq{Ks}, to a mean activity of \(\langle K \rangle / \tt = 4\ee^{-2s}\) in this limit.

\begin{figure*}
\begin{center}
\includegraphics[width=1\textwidth]{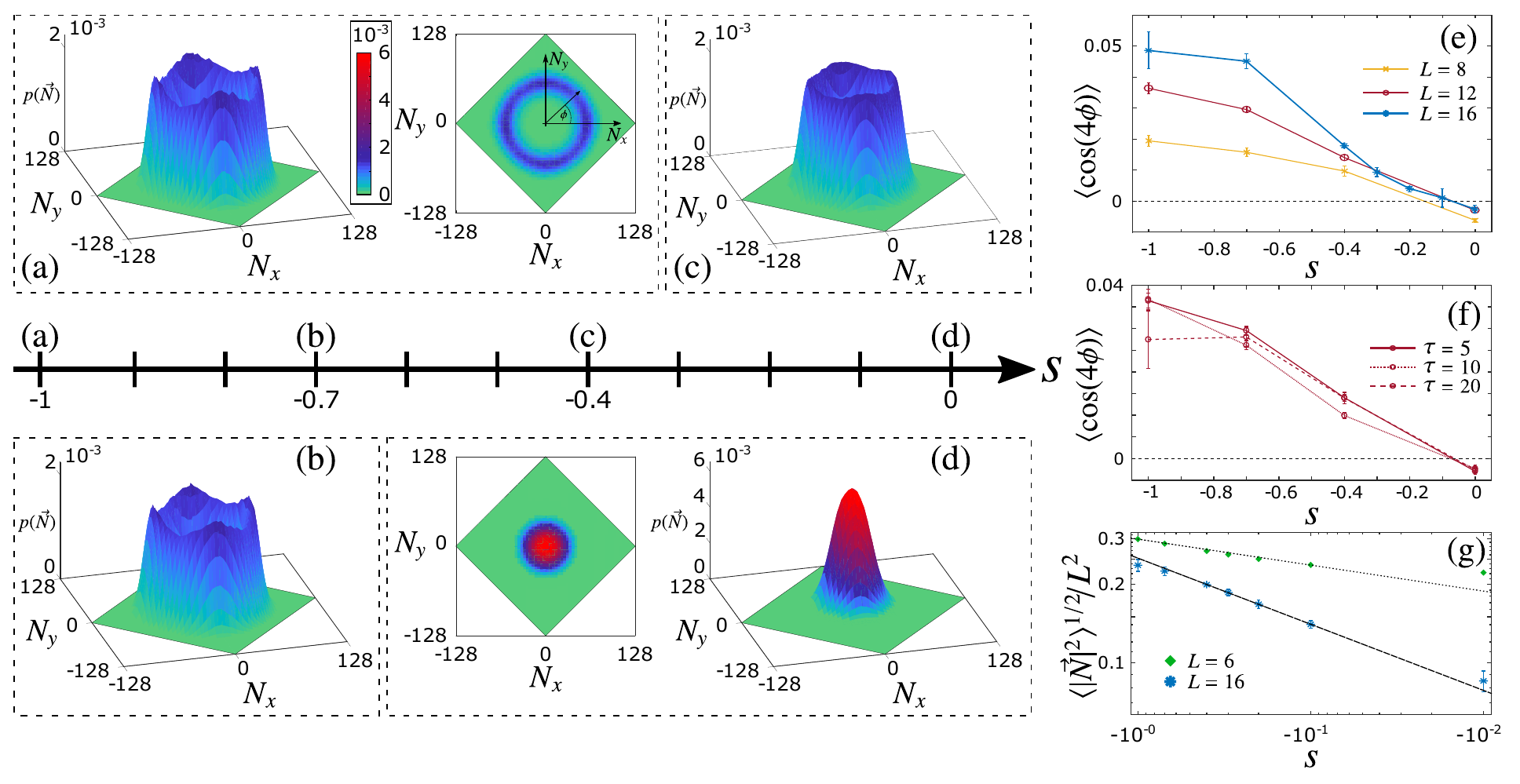}
\end{center}
\caption{
(a--d) Distribution \(p(\Nv)\) of magnetization \(\Nv\) in the quantum dimer model, for various values of the parameter \(s = \ln (v/t)\), as indicated on the central \(s\) axis, and system size \(L = 16\). (Note that the magnetization, like the flux \(\Phiv\), obeys \(\lvert N_x \rvert + \lvert N_y \rvert \le \frac{1}{2}L^2\).) For all \(s < 0\), the distribution has approximate circular symmetry, but with peaks along the square axes, corresponding to columnar order. The prominence of the peaks and the magnitude \(\lvert\Nv\rvert\) of the ring decrease as one approaches the RK point at \(s=0\), where the distribution is Gaussian around \(\Nv = \zerov\).
(e--f) Anisotropy measure \(\langle \cos (4\phi)\rangle\), where \(\tan \phi = N_y/N_x\), evaluated in the ground state \(\ket{\gs}\), versus \(s\). Panel (e) shows the dependence on system size \(L\), while in panel (f) the trajectory time \(\tt\) used in the simulations is varied with \(L = 12\) fixed. Positive values correspond to a distribution peaked along the square axes, confirming that the ordering is columnar and that it becomes more pronounced as \(\lvert s\rvert\) increases. (The small negative value at \(s=0\) is, we believe, a consequence of the discrete values taken by \(\Nv\).) 
(g) Root-mean-square magnetization magnitude \(\langle \lvert \Nv \rvert^2 \rangle^{1/2}\), corresponding roughly to the radius of the ring in the distribution of \(\Nv\), as a function of \(s\) for \(L = 6\) and \(16\). The dashed line shows a power-law fit to the data for \(L=16\) and \(-0.4 \le s \le -0.1\) with fitted exponent \(\beta\sub{eff} = 0.254\), while the dotted line shows an approximate fit to the \(L=6\) data, with \(\beta\sub{eff} = 0.1\).
}
\label{FigMagnetization}
\end{figure*}

Our main results regarding the phase structure of the QDM are displayed in \reffig{FigMagnetization}. Panels (a--d) show the ground-state order-parameter distribution,
\beq
p(\Nv') = \bra{\gs}\delta(\Nv - \Nv')\ket{\gs}\punc,
\eeq
where \(\Nv\) is the operator defined in \refeq{EqMagnetizationDefinition} and the expectation value in the QDM ground state \(\ket{\gs}\) is calculated as described in \refsec{SecConnectionToQDM}. For all negative \(s\) (\(v/\t < 1\)), the maxima of the distribution \(p(\Nv)\) occur for \(\Nv\) aligned with the square axes, indicating that the ground state has columnar order. Particularly for small \(\lvert s\rvert\), though, the selection of this order is weak, and the distribution has approximate \(\SO(2)\) symmetry under continuous rotations, with the largest probabilities occurring on a ring of fixed \(\lvert\Nv\rvert\).

To characterize quantitatively the degree of selection of columnar order, we consider the quantities \(\bra{\gs} \lvert\Nv\rvert^2\ket{\gs}\) and \(\bra{\gs} \cos (4\phi)\ket{\gs}\) where \(\tan\phi = N_y/N_x\). As panels (e--g) of \reffig{FigMagnetization} show, both of these quantities decrease as the RK point (\(s=0\)) is approached, in qualitative agreement with the results of Banerjee et al.\ \cite{Banerjee}.

The microscopic model is only symmetric under the discrete rotations of the lattice, and so the approximate \(\SO(2)\) symmetry of the order parameter \(\Nv\) is emergent. As argued by Fradkin et al.\ \cite{Fradkin2004}, this can be understood qualitatively by considering the renormalization group (RG) flow structure. Properties near the RK point are described by an effective action
\beq[EqAction]
\scS = \frac{1}{2}(\partial_0 h)^2 + \frac{1}{2}\rho_2(\del h)^2 + \frac{1}{2}\rho_4(\del^2 h)^2 + \lambda \cos(2\pi h) + \dotsb\punc,
\eeq
written in terms of a continuum height field \(h\), where \(\rho_2\), \(\rho_4\), and \(\lambda\) are real parameters, and \(\del\) and \(\partial_0\) denote the space and imaginary-time derivatives, respectively. The coefficient \(\rho_2 \sim \t - v \sim {-s}\) is tuned through zero at the RK point, which, in spite of the first-order nature of the phase transition, corresponds to a critical fixed point of the height field theory.

While the magnetization \(\Nv\) does not appear explicitly in \refeq{EqAction}, it is related to the coarse-grained height by
\beq[EqMagnetizationAndHeight]
N_x + \ii N_y \sim \exp\left[- \ii \frac{\pi}{2} \left(h + \frac{1}{2}\right)\right]\punc.
\eeq
The \(\lambda\) term therefore breaks \(\SO(2)\) down to the discrete subgroup of lattice symmetries, and determines the ultimate direction of the RG flow, towards columnar order for positive \(\lambda\). At the RG fixed point corresponding to the RK point, however, \(\lambda\) is strongly irrelevant, with RG eigenvalue \(y_\lambda = -12\). Standard scaling arguments in the presence of a dangerously irrelevant perturbation \cite{Senthil2004,Sreejith2014} therefore imply the existence of an additional length scale, \(\propto \lvert s\rvert^{-3}\) for small negative \(s\), with selection between columnar and plaquette order occurring only beyond this large scale \cite{Fradkin2004}. This is qualitatively consistent with the weak columnar ordering observed for small \(\lvert s \rvert\) at the system sizes accessible in our MC simulations.

The same effective action can be used to calculate the critical behavior of the the root-mean-square magnetization \(N\sub{rms} = \bra{\gs} \lvert\Nv\rvert^2\ket{\gs}^{1/2}\), shown in \reffig{FigMagnetization}(g). We find in \refapp{AppRMSN} that \(N\sub{rms} \sim L^2\lvert s\rvert^{1/2}\) for small negative \(s\) and large \(L\), corresponding to a critical exponent \(\beta = \frac{1}{2}\). Finite-size corrections, however, cause deviations from this scaling form and saturation at \(N\sub{rms} \sim L\sqrt{\ln L}\) when \(s = 0\). At fixed \(L\), crossover between these two forms, \(\sim s^{1/2}\) and \(\sim s^0\) (saturation), leads to a reduced effective exponent \(\beta\sub{eff}\). As shown in \reffig{FigMagnetization}(g), we find a reasonable fit to \(\beta\sub{eff} = 0.254\) for \(L = 16\) and \(\beta\sub{eff}\simeq 0.1\) using exact results for \(L = 6\), consistent with such a scenario. Further results at much larger system sizes would likely be needed to confirm the expected scaling behavior, and we leave this to future work.

Note that the unusual nature of the phase transition at \(s=0\), which is thermodynamically first-order but shows critical behavior on the negative-\(s\) side (\(v/\t < 1\)), is a common feature of constrained systems such as dimer models. A classic example is the Kasteleyn transition in 2D classical dimer models \cite{Kasteleyn1963}.

\section{Conclusions}

Our results here provide an example of the connection between the 
statistical properties of long-time trajectories of a classical system and the properties of the low-lying spectrum of a related quantum system. Here we have focused on the classical fully packed dimer model on the square lattice and, correspondingly, the quantum dimer model. The connection works both ways as we have illustrated: from the known existence of a quantum phase transition in the QDM at the RK point we infer the existence of a transition -- which we confirm numerically -- between active and inactive dynamical phases in the CDM. Conversely, from the statistics of atypical trajectories of the CDM we learn about the ground state properties of the QDM away from the RK point. Other examples of this classical--quantum connection include classical exclusion processes and {\it XXZ} chains \cite{Appert2008,Lecomte2012,Karevski2017}, and the one-dimensional Ising model with Glauber dynamics and the transverse field Ising chain \cite{Jack2010}.

For the QDM, our main result (see \reffig{FigMagnetization}) is that the ground state is the columnar phase for all \(v/\t < 1\). Our results in this regime also show an approximate emergent \(\SO(2)\) symmetry of the order parameter. Both of these findings agree with the observations of \refcite{Banerjee}, and contradict earlier results \cite{Syljuasen2006,Ralko2008}. We note that the method we use here is closer in spirit to that of the earlier work, based upon projector MC. For the CDM, the main result is the non-trivial structure of fluctuations in the dynamics away from typical behaviour. The first-order transition at $s=0$, see \reffig{FigActivity}(a), implies a coexistence in the equilibrium dynamics of space--time regions of high and low activity, and therefore a broad distribution of the dynamical order parameter, see \reffig{FigActivity}(b). Furthermore, the two competing phases display different kinds of structural order: while the inactive phase ($s>0$) is staggered, the active phase ($s<0$) is columnar (with both plaquette and mixed order being metastable due to their stability over shorter length scales). This change in the nature of configurations in order to optimise large dynamical fluctuations is reminiscent of what occurs in other systems, such as simple exclusion processes where -- even in a state where the typical activity and current are featureless -- rare inactive trajectories are associated with phase separated states and atypical large currents to hyperuniform (super-homogeneous) states \cite{Jack2015,Carollo2017,Karevski2017}. 

A consequence of the large fluctuations in the dynamics is that sampling rare trajectories is difficult. This is more so in a system like the CDM with periodic boundary conditions due to the constrained nature of configuration space and the conservation of the flux. To sample trajectory space we used transition path sampling [a Monte Carlo meta-dynamics in the space of trajectories guaranteed to converge to the $s$-ensemble \refeq{sens}], and to overcome the numerical difficulty of accessing exponentially suppressed trajectories we supplemented TPS with a version of  umbrella sampling in trajectory space \cite{Nemoto2016,Ray2017,Ray2017b,Klymko2017}. TPS is well suited to our problem as the CDM dynamics obeys detailed balance (and is in fact bi-stochastic). Our umbrella sampling could be improved by obtaining the reference dynamics in an adaptive manner, as is done in \cite{Nemoto2016} for cloning dynamics. Other interesting avenues to pursue include considering open boundary conditions (where we expect exploration of dynamics to be easier due to the absence of flux conservation), and to study in a similar manner as here dimer coverings in other geometries including higher dimensions.

The method we have presented can easily be generalized to other geometries and other systems. All that is required is that the system of interest has an RK point, i.e., that for certain values of the parameters the Hamiltonian is equivalent to a stochastic generator \cite{Castelnovo2005}. If that is the case, properties of the ground state away from the RK point can be recovered from the rare fluctuations of the stochastic system, just as we have done here for the QDM away from \(v=t\).

\acknowledgments The simulations used resources provided by the University of Nottingham High-Performance Computing Service. This work was supported by EPSRC Grant Nos.\ EP/M019691/1 (SP), EP/P034616/1 (CC \& AL), EP/K028960/1 (CC), and EP/M014266/1 (JPG).

\appendix
\section{$L=2$ CDM with open boundaries}
\label{app_open_2by2}

\begin{figure}
\includegraphics[width=\columnwidth]{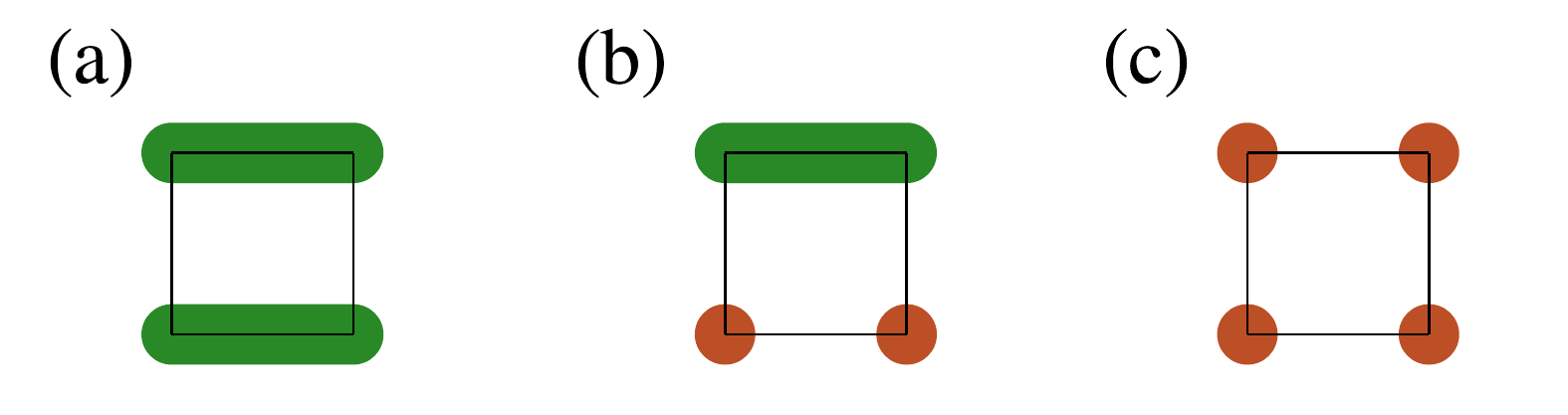}
\caption{
Dimer configurations of a \twobytwo\ region with open boundary conditions. There are seven configurations, which can be divided into three classes; one example of each is shown. (a) One of the two configurations with two dimers inside the included region. (b) One of the four configurations with a single dimer inside. (c) The single configuration with no dimers inside. In (b) and (c), the red circles represent sites whose dimers point out of the region.
}
\label{FigOpenTwoByTwo}
\end{figure}

We can obtain an approximation to the Doob-transformed dynamics (see \refsec{SecTPSUmbrella}) of a large system by focusing on the properties of a local region of size \twobytwo. This corresponds to a dimer model with four sites and {\em open boundary conditions}: while a dimer is connected to each of the sites, these dimers may be directed outwards and so not contained within the \twobytwo\ region. We can thus think of each site as occupied by a either a dimer or a monomer. As shown in \reffig{FigOpenTwoByTwo}, there are seven configurations in this open $L=2$ problem: two configurations with two dimers within the region, four configurations with one dimer and two monomers, and a single configuration with four monomers. The dynamics of the larger CDM, \refsec{SecCDM}, induces a dynamics between these seven states of the local \twobytwo\ region.

The dynamical generator in the reduced system has the form
\begin{widetext}
\begin{equation}
\W^{\twobytwo}  = \begin{bmatrix}
    -4b & b & b & b & b & 0 & 0\\
    b & -(a+b) & 0 & 0 & 0 & a & 0\\
    b & 0 & -(a+b) & 0 & 0 & 0 & a\\
    b & 0 & 0 & -(a+b) & 0 & 0 & a\\
    b & 0 & 0 & 0 & -(a+b) & a & 0\\
    0 & a & 0 & 0 & a & -(2a+c) & c\\
    0 & 0 & a & a & 0 & c & -(2a+c)\\
  \end{bmatrix}\punc,
\end{equation}
\end{widetext}
where the components correspond to configurations with, respectively, four monomers (first row), a single dimer and two monomers (rows 2--5), and two dimers (rows 6--7). The above generator has three kinds of transitions: between the two-dimer configurations at rate $c$, between single- and double-dimer configurations at rate $a$, and between the no-dimer and single-dimer configurations at rate $b$. The former kind of transition corresponds to a plaquette flip within the \twobytwo\ region, while the latter two are when the flip occurs at its boundary. The values of the rates depend on the size of the system on which the smaller region is embedded and can be obtained numerically from simulations. 

As explained in the main text, we can deform $\W^{\twobytwo}$ to obtain a SCGF for the number of flips from the largest eigenvalue of the deformed operator
\begin{widetext}
\begin{equation}
\W^{\twobytwo}_s  = \begin{bmatrix}
    -4b & b & b & b & b & 0 & 0\\
    b & -(a+b) & 0 & 0 & 0 & a & 0\\
    b & 0 & -(a+b) & 0 & 0 & 0 & a\\
    b & 0 & 0 & -(a+b) & 0 & 0 & a\\
    b & 0 & 0 & 0 & -(a+b) & a & 0\\
    0 & a & 0 & 0 & a & -(2a+c) & e^{-s} c\\
    0 & 0 & a & a & 0 & e^{-s} c & -(2a+c)\\
  \end{bmatrix}\punc.
  \label{W22s}
\end{equation}
\end{widetext}
We count only the transitions between the two-dimer configurations in the region, to avoid over-counting when we reconstruct the large system by overlaying \twobytwo\ regions, as the other transitions correspond to flips in neighbouring regions.

As an approximation to the Doob transform for the full \(L\times L\) system, we replace the exact vector $\bra{l_s}$ by the product of the left eigenvectors \(\bra{\ell}\) of \(\W^{\twobytwo}_s\) for each \twobytwo\ region. The components $\ell_c$ can be expressed in the form $\ell_c = e^{\varepsilon_c}$, where \(\varepsilon_c\) is a (dimensionless) ``energy'' associated to configuration \(c\). (Both \(\ell_c\) and \(\varepsilon_c\) depend on \(s\); we suppress this for clarity.)

We can characterize each configuration \(c\) by the number \(N_n\) of plaquettes with \(n\) dimers (i.e., the number in each class in \reffig{FigOpenTwoByTwo}); note that \(N_2 \equiv N\sub{f}(c)\), in the notation of \refsec{SecEffectiveReferenceDynamics}. The total number of plaquettes is \(N_0 + N_1 + N_2 = L^2\), and, the fact that the total number of dimers is constrained to be \(\frac{1}{2}L^2\) implies that \(2N_2 + N_1 = L^2\). Together these give \(N_0 = N_2 = \frac{1}{2}(L^2 - N_1)\) and allow us to express the dependence of the eigenvector on \(s\) (as well as on the rates \(a\), \(b\), and \(c\)) as $\varepsilon_s = D_s N_2$, using a single parameter \(D_s\) obtained from the diagonalisation of $\W^{\twobytwo}_s$.

This value of \(D_s\) specifies a dynamics that generates the exact \(s\)-ensemble for the open \twobytwo\ problem and that we use as a starting point when optimizing the reference dynamics for the full lattice (see \refsec{SecOptimizationOfReferenceDynamics}).

\section{Root-mean-square magnetization near RK point}
\label{AppRMSN}

As argued in \refsec{SecResults}, close to the RK point and below the length scale for columnar ordering, one can set \(\lambda = 0\) and drop higher-order terms in \refeq{EqAction}, leaving a quadratic action \(\scS\). Expressing the magnetization \(\Nv\) in terms of the height \(h\) using \refeq{EqMagnetizationAndHeight}, we then have
\beq[EqRMSN]
N\sub{rms}^2 \sim L^2 \int \dd^2 \rv \, \ee^{-\frac{\pi^2}{8} h\sub{rms}^2(\rv)}\punc,
\eeq
where \(h\sub{rms}^2(\rv) = \bra{\gs} \left[h(\rv) - h(\boldsymbol{0})\right]^2 \ket{\gs}\) is the mean-square height difference for positions separated by \(\rv\).

By writing \(\scS\) in terms of the Fourier transform of \(h\), one can express this ground-state expectation value as an integral over wavevector \(\kv\) and frequency \(\omega\). Integrating over \(\omega\) and the angle between \(\kv\) and \(\rv\) gives
\beq[EqhrmsIntegral]
h\sub{rms}^2(\rv) = \frac{1}{2\pi}\int_0^\Lambda \dd k \frac{1-J_0(k \lvert \rv \rvert)}{\sqrt{\rho_2 + \rho_4 k^2}}\punc,
\eeq
where \(k = \lvert \kv \rvert\), \(\Lambda\) is an ultraviolet cutoff of order the inverse of the lattice spacing, and \(J_0\) is a Bessel function.

For \(\lvert\rv\rvert \gg \Lambda^{-1}\), the integral in \refeq{EqhrmsIntegral} can be evaluated analytically, giving
\beq
N\sub{rms}^2 \sim L^2 \left[\Psi\left(L\sqrt{\rho_2/\rho_4}\right) - \Psi\left(\Lambda^{-1}\sqrt{\rho_2/\rho_4}\right)\right]\punc,
\eeq
in terms of the function
\beq
\Psi(x) = \int_1^{x/2}\dd u \, u \ee^{2I_0(u)K_0(u)}\punc,
\eeq
where \(I_0\) and \(K_0\) are modified Bessel functions of the first and second kind, respectively. The behavior for smaller \(\lvert\rv\rvert\) (i.e., of order the lattice spacing) is not well described by the continuum action, and, according to \refeq{EqRMSN}, will make a contribution of order \(L^2\Lambda^2\) to \(N\sub{rms}^2\).

The results quoted in \refsec{SecResults} for the root-mean-square magnetization \(N\sub{rms}\) follow from the asymptotic behavior of the function \(\Psi\). For large \(x\), \(\Psi(x) \sim x^2\), and so in the large-\(L\) limit with fixed nonzero \(\rho_2 \sim \lvert s\rvert\), \(N\sub{rms}\sim L^2\lvert s\rvert^{1/2}\). (The magnetization is therefore extensive, as expected in an ordered phase.) For small \(x\), \(\Psi(x) \sim \lvert\ln x\rvert\), and so \(N\sub{rms} \sim L\sqrt{\ln L}\) at \(s = 0\). In both cases, the contributions from lattice-scale effects are of lower order.


\begin{thebibliography}{99}
\bibitem{FootnoteProductState} The product state in \refeq{EqProductState} has \(\bra{\theta} N\sub{f} \ket{\theta} = \frac{1}{4}L^2(1 + \sin^4\theta + \cos^4\theta)\).
\end{thebibliography}
\end{document}